\documentclass[11pt,twoside]{article}
\usepackage{amsmath}
\usepackage{amssymb}
\usepackage{amscd}
\usepackage{mathrsfs}
\usepackage{graphicx}
\usepackage{epstopdf}
\renewcommand{\a}{\alpha}
\renewcommand{\b}{\beta}
\newcommand{\g}{\gamma}
\newcommand{\G}{\Gamma}
\renewcommand{\d}{\delta}
\newcommand{\e}{\varepsilon}
\newcommand{\z}{{\zeta}}
\renewcommand{\th}{\theta}
\newcommand{\Th}{\Theta}
\renewcommand{\k}{\kappa}
\renewcommand{\l}{\lambda}

\newcommand{\m}{\mu}
\newcommand{\n}{\nu}

\renewcommand{\r}{\rho}
\newcommand{\s}{\sigma}
\renewcommand{\t}{\tau}
\newcommand{\om}{{\omega}}
\newcommand{\Cs}{{\rlap{\lower3pt\hbox{\textnormal{\LARGE \char'040}}}{\Gamma}}{}}

\newcommand{\QI}{\Lambda}
\newcommand{\Eo}{{\scriptstyle{\mathrm{E}}}}

\newcommand{\oh}{\tfrac{1}{2}}

\newcommand{\osq}{\tfrac{1}{\surd2}}

\newcommand{\cj}[1]{\overline{#1}}
\newcommand{\lin}{{\scriptscriptstyle\bigstar}}
\renewcommand{\.}{{\scriptstyle\boldsymbol{\dot{}}}}
\newcommand{\td}{\tilde}
\newcommand{\cev}[1]{\overleftarrow{#1}}
\newcommand{\Lll}{{\scriptscriptstyle{\mathrm{L}}}}
\newcommand{\Rrr}{{\scriptscriptstyle{\mathrm{R}}}}
\newcommand{\interaction}{{}_{\sst{\mathrm{int}}}}
\newcommand{\sbot}{{\scriptscriptstyle\bot}}
\newcommand{\sbo}{{\!\sbot}}
\newcommand{\spar}{{\scriptscriptstyle\|}}
\newcommand{\fl}{\flat}
\newcommand{\bk}{{\bar\kappa}}
\newcommand{\bl}{{\bar\lambda}}
\newcommand{\bm}{{\bar\mu}}
\newcommand{\bs}{{\bar s}}
\newcommand{\bu}{{\bar u}}
\newcommand{\bv}{{\bar v}}
\newcommand{\bw}{{\bar w}}
\newcommand{\bvv}{{\bar\vv}}
\newcommand{\be}{{\bar\varepsilon}}
\newcommand{\bze}{{\bar\zeta}}
\newcommand{\B}{{\boldsymbol{B}}}
\newcommand{\Bb}{{\scriptscriptstyle{\boldsymbol{B}}}}
\newcommand{\E}{{\boldsymbol{E}}}
\newcommand{\F}{{\boldsymbol{F}}}
\newcommand{\Fl}{\F^\lin}
\newcommand{\Fc}{\cj{\F}}
\newcommand{\Fa}{\cj{\F}{}^\lin}
\newcommand{\Gb}{{\boldsymbol{G}}}
\newcommand{\GA}{{\boldsymbol{\Gamma}}}
\renewcommand{\H}{{\boldsymbol{H}}}
\newcommand{\I}{{\boldsymbol{I}}}
\newcommand{\M}{{\boldsymbol{M}}}
\newcommand{\Mm}{{\scriptscriptstyle{\boldsymbol{M}}}}
\renewcommand{\P}{{\boldsymbol{P}}}
\newcommand{\Pm}{\P_{\!\!m}}
\newcommand{\Px}[1]{\P_{\!\!#1}}
\newcommand{\Pz}{\P_{{\!\!}_0}}
\let\Sec=\S
\renewcommand{\S}{{\boldsymbol{S}}}
\newcommand{\Sc}{\cj{\S}}
\newcommand{\Sa}{\cj{\S}{}^\lin}
\newcommand{\Sl}{\S{}^\lin}
\newcommand{\T}{{\boldsymbol{T}}}
\newcommand{\Tt}{{\scriptscriptstyle{\boldsymbol{T}}}}
\newcommand{\U}{{\boldsymbol{U}}}
\newcommand{\Uc}{\cj{\U}}
\newcommand{\Ua}{\cj{\U}{}^\lin}
\newcommand{\Ul}{\U{}^\lin}
\newcommand{\V}{{\boldsymbol{V}}}
\newcommand{\Vc}{\cj{\V}}
\newcommand{\W}{{\boldsymbol{W}}}
\newcommand{\Wc}{\cj{\W}}
\newcommand{\Wa}{\cj{\W}{}^\lin}
\newcommand{\Wl}{\W{}^\lin}
\newcommand{\X}{{\boldsymbol{X}}}
\newcommand{\Y}{{\boldsymbol{Y}}}
\newcommand{\Yc}{\cj{\Y}}
\newcommand{\Ya}{\cj{\Y}{}^\lin}
\newcommand{\Yl}{\Y{}^\lin}
\newcommand{\Z}{{\boldsymbol{Z}}}

\newcommand{\Zl}{\Z{}^\lin}
\newcommand{\Lie}{\mathfrak{L}}
\newcommand{\Ug}{\boldsymbol{\mathrm{U}}}
\newcommand{\CC}{{\mathbb{C}}}
\newcommand{\LL}{{\mathbb{L}}}
\newcommand{\NN}{{\mathbb{N}}}
\newcommand{\RR}{{\mathbb{R}}}
\newcommand{\VV}{{\mathbb{V}}}
\newcommand{\ZZ}{{\mathbb{Z}}}
\newcommand{\RRr}{{\scriptscriptstyle{\mathbb{R}}}}
\newcommand{\Dcal}{{\mathcal{D}}}
\newcommand{\Hcal}{{\mathcal{H}}}
\newcommand{\Lcal}{{\mathcal{L}}}
\newcommand{\Ocal}{{\mathcal{O}}}
\newcommand{\Qcal}{{\mathcal{Q}}}
\newcommand{\Scal}{{\mathcal{S}}}
\newcommand{\Ucal}{{\mathcal{U}}}
\newcommand{\Vcal}{{\mathcal{V}}}
\newcommand{\Xcal}{{\mathcal{X}}}
\newcommand{\Ycal}{{\mathcal{Y}}}
\newcommand{\Zcal}{{\mathcal{Z}}}
\newcommand{\Hfr}{\mathfrak{H}}
\newcommand{\lfr}{\mathfrak{l}}

\newcommand{\DC}{{\boldsymbol{\Dcal}}}
\newcommand{\HC}{{\boldsymbol{\Hcal}}}
\newcommand{\LC}{{\boldsymbol{\Lcal}}}
\newcommand{\OC}{{\boldsymbol{\Ocal}}}
\newcommand{\QC}{{\boldsymbol{\Qcal}}}

\newcommand{\VC}{{\boldsymbol{\Vcal}}}
\newcommand{\XC}{{\boldsymbol{\Xcal}}}
\newcommand{\YC}{{\boldsymbol{\Ycal}}}
\newcommand{\ZC}{{\boldsymbol{\Zcal}}}
\newcommand{\DCo}{\DC_{\!\circ}}
\newcommand{\DCh}{{\rlap{\;/}\DC}}
\newcommand{\uDCh}{{\rlap{\;/}{\ul\DC}}}
\newcommand{\DCho}{{\rlap{\;/}\DCo}}
\newcommand{\End}{\operatorname{End}}
\newcommand{\Aut}{\operatorname{Aut}}
\newcommand{\Ker}{\operatorname{Ker}}
\newcommand{\Id}[1]{{1\!\!1}\!{}_{#1}{}}
\newcommand{\id}{{1\!\!1}}
\newcommand{\dO}{\mathrm{d}}

\newcommand{\DO}{\mathrm{D}}
\newcommand{\HO}{\mathrm{H}}
\newcommand{\JO}{\mathrm{J}}
\newcommand{\kO}{\mathrm{k}}
\newcommand{\TO}{\mathrm{T}}
\newcommand{\TS}{\TO^{*}\!}
\newcommand{\VO}{\mathrm{V}}

\newcommand{\dt}{\dO\tt}

\newcommand{\eO}{\mathrm{e}}
\newcommand{\iO}{\mathrm{i}}
\newcommand{\ten}[1]{\operatorname*{\otimes}_{\!{\scriptscriptstyle #1}} }
\newcommand{\dir}[1]{\operatorname*{\oplus}_{\!{\scriptscriptstyle #1}} }
\def\cart#1{\mathop\times\limits_{\scriptscriptstyle #1}}
\newcommand{\we}{{\,\wedge\,}}
\newcommand{\weu}[1]{{\wedge^{\!#1}}}
\newcommand{\mdots}{{\cdot}{\cdot}{\cdot}}
\newcommand{\pint}{\mathord{\rfloor}}
\newcommand{\comp}{\mathbin{\raisebox{1pt}{$\scriptstyle\circ$}}}
\newcommand{\tn}{{\,\otimes\,}}
\newcommand{\bang}[1]{{\langle#1\rangle}}

\newcommand{\ket}[1]{{|#1\rangle}}
\newcommand{\Ii}[2]{{}^{\smash{#1}}_{\phantom{\smash{#1}}\!\smash{#2}}}
\newcommand{\iI}[2]{{}_{\smash{#1}}^{\phantom{\smash{#1}}\!\smash{#2}}}
\newcommand{\iIi}[3]{{}_{\smash{#1}\phantom{\smash{#2}}\!\!\smash{#3}}^{\phantom{\smash{#1}}\!\smash{#2}}}
\newcommand{\si}[2]{\sigma\iI{#1}{#2}}
\newcommand{\sA}{{\scriptscriptstyle A}}
\newcommand{\sB}{{\scriptscriptstyle B}}
\newcommand{\sQ}{{\scriptscriptstyle Q}}
\newcommand{\cA}{{\sA\.}}
\newcommand{\AAd}{{\sA\cA}}
\newcommand{\zeA}{{\zeta_\sA}}
\newcommand{\bzeA}{{\bze_\cA}}
\newcommand{\Asf}{{\mathsf{A}}}
\newcommand{\Bsf}{{\mathsf{B}}}
\newcommand{\bb}{{\mathsf{b}}}
\newcommand{\Csf}{{\mathsf{C}}}
\newcommand{\Xsf}{{\mathsf{X}}}
\renewcommand{\tt}{{\mathsf{t}}}
\newcommand{\uu}{{\mathsf{u}}}
\newcommand{\vv}{{\mathsf{v}}}

\newcommand{\yy}{{\mathsf{y}}}
\newcommand{\sref}[1]{\Sec\ref{#1}}
\newcommand{\ie}{i.e$.$}
\newcommand{\sst}{\scriptscriptstyle}
\newcommand{\onto}{\rightarrowtail}
\newcommand{\ul}{\underline}
\newcommand{\YL}{\Y\!_{\!\Lll}} \newcommand{\YR}{\Y\!_{\!\Rrr}}
\newcommand{\YLc}{\Yc\!_{\!\Lll}} \newcommand{\YRc}{\Yc\!_{\!\Rrr}}
\newcommand{\FL}{\F_{\!\!\Lll}} \newcommand{\FR}{\F_{\!\!\Rrr}}
\newcommand{\FLc}{\Fc_{\!\!\Lll}} \newcommand{\FRc}{\Fc_{\!\!\Rrr}}
\newcommand{\FLl}{\Fl_{\!\!\Lll}} \newcommand{\FRl}{\Fl_{\!\!\Rrr}}
\newcommand{\der}{\delta^{\sst{(r)}}}
\newcommand{\ellr}{\ell^{\sst{(r)}}}
\newcommand{\uLar}{\ul\Lambda^{\!\sst{(r)}}}
\newcommand{\Hivac}{{\scriptstyle\Hcal}_{\sst0}}
\newcommand{\bHivac}{\bar{\scriptstyle\Hcal}_{\sst0}}
\newcommand{\gh}{\omega}
\newcommand{\ghR}{\gh_\Rrr} \newcommand{\ghL}{\gh_\Lll}
\newcommand{\agh}{\varpi}
\newcommand{\aghR}{\agh_\Rrr} \newcommand{\aghL}{\agh_\Lll}
\newcommand{\FLa}{\Fa_{\!\!\Lll}} \newcommand{\FRa}{\Fa_{\!\!\Rrr}}
\newcommand{\QIr}{\QI^{\!\sst{(r)}}}
\newcommand{\Pmm}{\P\!\!_{{\scriptstyle\times}}}
\newcommand{\scc}{{\mathsf{c}}}
\newcommand{\sco}[3]{\scc\Ii{#1}{#2#3}}
\newcommand{\bwe}{\,{\barwedge}\,}
\newsavebox{\boxLcqs}
\savebox{\boxLcqs}[20pt]{
\begin{picture}(20,20)(0,5)
\put(10,20){\vector(-1,-2){7}}
\put(5,10){\line(-1,-2){5}}
\put(20,0){\vector(-1,2){7}}
\put(15,10){\line(-1,2){5}}
\linethickness{.2pt}
\qbezier(10.,20.)(10.5,19.)(10.,18.)
\qbezier(10.,18.)(9.,17.)(10.,16.)
\qbezier(10.,16.)(11.,15.)(10.,14.)
\qbezier(10.,14.)(9.,13.)(10.,12.)
\qbezier(10.,12.)(11.,11.)(10.,10.)
\qbezier(10.,10.)(9.,9.)(10.,8.)
\qbezier(10.,8.)(11.,7.)(10.,6.)
\qbezier(10.,6.)(9.,5.)(10.,4.)
\qbezier(10.,4.)(11.,3.)(10.,2.)
\qbezier(10.,2.)(9.,1.)(10.,0)
\end{picture} }
\newcommand{\Lcqs}{\usebox{\boxLcqs}}
\newsavebox{\boxLSqs}
\savebox{\boxLSqs}[20pt]{
\begin{picture}(20,20)(0,5)
\put(10,10){\vector(-1,1){7}}
\put(10,10){\line(-1,1){10}}
\put(20,0){\vector(-1,1){7}}
\put(20,0){\line(-1,1){10}}
\linethickness{.2pt}
\qbezier(10.,0)(9.,1.)(10.,2.)
\qbezier(10.,2.)(11.,3.)(10.,4.)
\qbezier(10.,4.)(9.,5.)(10.,6.)
\qbezier(10.,6.)(11.,7.)(10.,8.)
\qbezier(10.,8.)(9.,9.)(10.,10.)
\end{picture} }
\newcommand{\LSqs}{\usebox{\boxLSqs}}
\newsavebox{\boxLcQs}
\savebox{\boxLcQs}[20pt]{
\begin{picture}(20,20)(0,5)
\put(10,10){\vector(-1,-1){7}}
\put(10,10){\line(-1,-1){10}}
\put(20,0){\vector(-1,1){7}}
\put(20,0){\line(-1,1){10}}
\linethickness{.2pt}
\qbezier(10.,20.)(11.,19.)(10.,18.)
\qbezier(10.,18.)(9.,17.)(10.,16.)
\qbezier(10.,16.)(11.,15.)(10.,14.)
\qbezier(10.,14.)(9.,13.)(10.,12.)
\qbezier(10.,12.)(11.,11.)(10.,10.)
\end{picture} }
\newcommand{\LcQs}{\usebox{\boxLcQs}}
\newsavebox{\boxLCqc}
\savebox{\boxLCqc}[20pt]{
\begin{picture}(20,20)(0,5)
\put(0,20){\vector(1,-1){7}}
\put(0,20){\line(1,-1){10}}
\put(10,10){\vector(1,-1){7}}
\put(10,10){\line(1,-1){10}}
\linethickness{.2pt}
\qbezier(10.,0)(9.,1.)(10.,2.)
\qbezier(10.,2.)(11.,3.)(10.,4.)
\qbezier(10.,4.)(9.,5.)(10.,6.)
\qbezier(10.,6.)(11.,7.)(10.,8.)
\qbezier(10.,8.)(9.,9.)(10.,10.)
\end{picture} }
\newcommand{\LCqc}{\usebox{\boxLCqc}}
\newsavebox{\boxLSQs}
\savebox{\boxLSQs}[20pt]{
\begin{picture}(20,20)(0,5)
\put(20,0){\vector(-1,1){7}}
\put(20,0){\line(-1,1){10}}
\put(10,10){\vector(-1,1){7}}
\put(10,10){\line(-1,1){10}}
\linethickness{.2pt}
\qbezier(10.,20.)(11.,19.)(10.,18.)
\qbezier(10.,18.)(9.,17.)(10.,16.)
\qbezier(10.,16.)(11.,15.)(10.,14.)
\qbezier(10.,14.)(9.,13.)(10.,12.)
\qbezier(10.,12.)(11.,11.)(10.,10.)
\end{picture} }
\newcommand{\LSQs}{\usebox{\boxLSQs}}
\newsavebox{\boxLSqC}
\savebox{\boxLSqC}[20pt]{
\begin{picture}(20,20)(0,5)
\put(10,10){\vector(-1,1){7}}
\put(10,10){\line(-1,1){10}}
\put(20,20){\vector(-1,-1){7}}
\put(20,20){\line(-1,-1){10}}
\linethickness{.2pt}
\qbezier(10.,0)(9.,1.)(10.,2.)
\qbezier(10.,2.)(11.,3.)(10.,4.)
\qbezier(10.,4.)(9.,5.)(10.,6.)
\qbezier(10.,6.)(11.,7.)(10.,8.)
\qbezier(10.,8.)(9.,9.)(10.,10.)
\end{picture} }
\newcommand{\LSqC}{\usebox{\boxLSqC}}
\newsavebox{\boxLCQc}
\savebox{\boxLCQc}[20pt]{
\begin{picture}(20,20)(0,5)
\put(0,20){\vector(1,-1){7}}
\put(0,20){\line(1,-1){10}}
\put(10,10){\vector(1,-1){7}}
\put(10,10){\line(1,-1){10}}
\linethickness{.2pt}
\qbezier(10.,20.)(11.,19.)(10.,18.)
\qbezier(10.,18.)(9.,17.)(10.,16.)
\qbezier(10.,16.)(11.,15.)(10.,14.)
\qbezier(10.,14.)(9.,13.)(10.,12.)
\qbezier(10.,12.)(11.,11.)(10.,10.)
\end{picture} }
\newcommand{\LCQc}{\usebox{\boxLCQc}}
\newsavebox{\boxLCQS}
\savebox{\boxLCQS}[20pt]{
\begin{picture}(20,20)(0,5)
\put(0,20){\vector(1,-2){7}}
\put(5,10){\line(1,-2){5}}
\put(10,0){\vector(1,2){7}}
\put(15,10){\line(1,2){5}}
\linethickness{.2pt}
\qbezier(10.,20.)(11,19.)(10.,18.)
\qbezier(10.,18.)(9.,17.)(10.,16.)
\qbezier(10.,16.)(11.,15.)(10.,14.)
\qbezier(10.,14.)(9.,13.)(10.,12.)
\qbezier(10.,12.)(11.,11.)(10.,10.)
\qbezier(10.,10.)(9.,9.)(10.,8.)
\qbezier(10.,8.)(11.,7.)(10.,6.)
\qbezier(10.,6.)(9.,5.)(10.,4.)
\qbezier(10.,4.)(10.5,3.)(10.,2.)
\qbezier(10.,2.)(9.,1.)(10.,0)
\end{picture} }
\newcommand{\LCQS}{\usebox{\boxLCQS}}
\title{Two-spinor geometry and gauge freedom}
\date{{\small March 6, 2014\\ (July 1, 2014)} }
\author{Daniel Canarutto\\[6pt]
{\small\it Dipartimento di Matematica e Informatica ``U.~Dini'', }\\
{\small\it Via S. Marta 3, 50139 Firenze, Italia}\\
{\small email:~daniel.canarutto@unifi.it}\\
{\small http://www.dma.unifi.it/\char126 canarutto}}
\begin{document}
\bibliographystyle{alpha}
\maketitle
\begin{center}
{\emph{Dedicated to Luigi Mangiarotti and Marco Modugno}}
\end{center}

\begin{abstract}\noindent
Gauge freedom in quantum particle physics is shown to arise
in a natural way from the geometry of two-spinors (Weyl spinors).
Various related mathematical notions are reviewed,
and a special ansatz of the kind ``the system defines the geometry''
is discussed in connection with the stated results.
\end{abstract}

\bigbreak

\bigbreak

\tableofcontents

\thispagestyle{empty}
\vfill\newpage
\thispagestyle{empty}

\section*{Introduction}

\begin{flushright}
{\sc entia non svnt mvltiplicanda praeter necessitatem}\par
{\emph{(Entities are not to be multiplied beyond necessity)}}\par
William of Ockam
\end{flushright}

\bigbreak\noindent
In previous papers I discussed some partly original notions
related to quantum particle physics,
including a ``minimal geometric data''\footnote{
This locution was crafted by A.~Jadczyk,
with whom I collaborated in these topics.} 
approach to in 2-spinor geometry and field theories%
~\cite{CJ97a,CJ97b,C98,C00b,C07,C10a},
the geometry of distributional bundles in terms of Fr\"olicher smoothness%
~\cite{C04a}
and its application to quantum bundles, quantum connections
and particle interactions~\cite{C05,C09a,C12a}.
While these mathematical results have been offered as they are,
and perhaps can be read as small bits of clarification
about a rather confused matter,
they are actually driven by a somewhat radical ansatz,
related in part to ideas once proposed by Penrose~\cite{Pe71},
which was exposed in an essay~\cite{C11a}
presented for the 2011 contest of the Foundational Questions Institute.

Section~\ref{s:Two-spinors and gauge field theory}
is devoted to a sketch of 2-spinor geometry
and the treatment of gauge field theories based on it,
together with a somewhat novel discussion of symmetry breaking.
The notions of quantum bundles, quantum states and quantum interactions
are reviewed in section~\ref{s:Quantum states and interactions}.
The paper's main section is~\sref{s:Gauge freedom},
where the notion of gauge freedom
is looked at from different points of view, and, in particular,
in terms of 2-spinor geometry.
Finally, some foundational issues related to the exposed ideas are discussed;
there I do not claim to have demonstrated my main thesis,
but argue that some clues support it.

\section{Two-spinors and gauge field theory}
\label{s:Two-spinors and gauge field theory}
\subsection{Two-spinor basics}
\label{ss:Two-spinor basics}

If $\V$ is a complex vector space
then Hermitian transposition is a natural anti-linear involution of $\V\tn\Vc$
(where $\Vc$ denotes the conjugate space),
determining a decomposition into the direct sum
of the \emph{real} eigenspaces
corresponding to eigenvalues $\pm1$\,, namely
$$\V\tn\Vc=\HO(\V\tn\Vc)\oplus \iO\,\HO(\V\tn\Vc)~,$$
called the \emph{Hermitian} and \emph{anti-Hermitian} subspaces, respectively

Starting from a $2$-dimensional complex vector space $\S$,
with no further assumption,
the above basic construction gives rise to a rich algebraic structure:
\smallbreak\noindent
$\bullet~$%
The Hermitian subspace of \hbox{$\weu2\S\tn\weu2\Sc$}
is a real $1$-dimensional vector space with a distinguished orientation;
its positively oriented semispace $\LL^2$
(whose elements are of the type $w\tn\bw$\,, $w\in\weu2\S$)
has the square root semispace $\LL$,
which will can be identified with the space of \emph{length units}~%
\cite{JMV10,C12a}.
\smallbreak\noindent
$\bullet~$%
The $2$-spinor space is defined to be $\U:=\LL^{-1/2}\tn\S$.
The space $\weu2\U$ is naturally endowed with a Hermitian metric,
namely the identity element in
$$\HO[(\weu{2}\Ua)\tn(\weu{2}\Ul)]
\cong\LL^2\tn\HO[(\weu{2}\Sa)\tn(\weu{2}\Sl)]~,$$
so that normalised `symplectic forms' $\e\in\weu2\Ul$
constitute a $\Ug(1)$-space (any two of them are related by a phase factor).
Each $\e$ yields the isomorphism
\hbox{$\e^\fl:\U\to\Ul:u\mapsto u^\fl:=\e(u,\_)$}\,.
\smallbreak\noindent
$\bullet~$%
The identity element in $\HO[(\weu{2}\Ua)\tn(\weu{2}\Ul)]$
can be written as $\e\tn\be$ where $\e\in\weu2\Ul$
is any normalised element.
This natural object can also be seen as a bilinear form $g$ on $\U\tn\Uc$,
via the rule
\hbox{$g(u\tn\bv,r\tn\bs)=\e(u,r)\,\be(\bv,\bs)$}
extended by linearity.
Its restriction to the Hermitian subspace $\H\equiv\HO(\U\tn\Uc)$
turns out to be a Lorentz metric.
Null elements in $\H$ are of the form $\pm u\tn\bu$ with $u\in\U$
(thus there is a distinguished time-orientation in $\H$).
\smallbreak\noindent
$\bullet~$%
Let $\W\equiv\U\oplus\Ua$.
The linear map
$\g:\U\tn\Uc\to\End(\W):y\mapsto\g[y]$ acting as
$$\td\g(r\tn\bs)[u,\bl]
=\sqrt2\bigl(\bang{\bl,\bs}\,p\,,\bang{r^\fl,u}\,\bs^\fl\,\bigr)$$
is well-defined independently of the choice of the normalised $\e\in\weu2\Ul$
yielding the isomorphism $\e^\fl$.
Its restriction to $\H$ turns out to be a Clifford map.
Thus one is led to regard $\W\equiv\U\oplus\Ua$ as the space of Dirac spinors,
decomposed into its Weyl subspaces.
The anti-isomorphism $\W\to\Wl:(u,\bl)\mapsto(\l,\bu)$
is the usual \emph{Dirac adjunction}
(\hbox{$\psi\mapsto\bar\psi$} in traditional notation),
associated with a Hermitian product having the signature $({+},{+},{-},{-})$\,.

\smallbreak
An arbitrary basis $(\xi_\sA)$ of $\S$, $\scriptstyle{A}=1,2$\,,
determines bases of the various associated spaces,
in particular the bases $l\in\LL$ (a length unit),
$\bigl(\zeA\bigr)\equiv\bigl(l^{-1/2}\,\xi_\sA \bigr)\subset\U$,
$\e\in\weu2\Ul$.
We have $\e=\e_{\sA\sB}\,\z^\sA\we\z^\sB$\,,
where $\bigl(\z^\sA\bigr)\subset\Ul$ is the dual basis of $\bigl(\zeA\bigr)$
and $\bigl(\e_{\sA\sB}\bigr)$
denotes the antisymmetric Ricci matrix.
As for the basis of $\H\equiv\HO(\U\tn\Uc)$ associated with $(\zeA)$
one usually considers the \emph{Pauli basis} $\bigl(\t_\l\bigr)$\,,
given by
$\t_\l\equiv\osq\,\si{\l}{\AAd}\,\zeA\tn\bzeA$
where $(\si{\l}{\AAd})$\,, $\l=0,1,2,3$\,, denotes the $\l$-th Pauli matrix
(dotted indices refer to components in conjugate spaces).
This basis is readily seen to be $g$-orthonormal.
The associated \emph{Weyl basis} of $\W$
is defined to be the basis $(\z_\a)$, $\a=1,2,3,4$, given by
$$(\z_1\,,\z_2\,,\z_3,\z_4):=(\z_1\,,\z_2\,,-\bze^1,-\bze^2)~,$$
where $\z_1$ is a simplified notation for $(\z_1\,,0)$, and the like.

\smallbreak\noindent{\bf Remark.}~%
In contrast to the usual $2$-spinor formalism,
no symplectic form is fixed.
The  $2$-form $\e$ is unique up to a phase factor
which depends on the chosen 2-spinor basis,
and determines isomorphisms
$\e^\fl:\U\to\Ul$ and $\e^\#:\Ul\to\U$\,.
Also note that no Hermitian form on $\S$ or $\U$ is assigned;
actually, because of the Lorentz structure of $\H$\,,
the choice of such an object turns out to be equivalent
to the choice of an `observer'.
\smallbreak

We now consider a complex vector bundle $\S\onto\M$
with $2$-dimensional fibers.
By performing the above sketched constructions fiberwise
we obtain various vector bundles, which are denoted, for simplicity,
by the corresponding symbols.
We observe that some appropriate topological restrictions
are implicit in what follows;
we'll assume the needed hypotheses to hold without further comment.

A linear connection $\Cs$ on $\S$ determines linear connections
on the associated bundles, and, in particular, connections
$G$ of $\LL$, $Y$ of $\weu2\U$ and $\td\G$ of $\H$;
on turn, it can be expressed in terms of these as
$$\Cs\iIi{a}{\sA}{\sB}=(G_a+\iO\,Y_a)\d\Ii{\sA}{\sB}
+\oh\,\td\G\iIi{a}{\sA\cA}{\sB\cA}~.$$

If $\M$ is $4$-dimensional, then a \emph{tetrad}
is defined to be a linear morphism $\Th:\TO\M\to\LL\tn\H$.
An invertible tetrad determines, by pull-back,
a Lorentz metric on $\M$ and a metric connection of $\TO\M\onto\M$,
as well as a Dirac morphism $\TO\M\to\LL\tn\End\W$.

A non-singular field theory in the above geometric environment
can be naturally formulated~\cite{C98}
even if $\Th$ is not required to be invertible everywhere.
If the invertibility requirement is satisfied
then one gets essentially the standard Einstein-Cartan-Maxwell-Dirac theory,
but with some redefinition of the fundamental fields:
these are now the $2$-spinor connection $\Cs$,
the tetrad $\Th$,
the Maxwell field $F$ and the Dirac field $\psi:\M\to\LL^{-3/2}\tn\W$.
Gravitation is represented by $\Th$ and $\td\G$ together.
$G$ is assumed to have vanishing curvature,
$\dO G=0$, so that we can find local charts such that $G_a=0$\,;
this amounts to `gauging away' the conformal `dilaton' symmetry.
Coupling constants arise as covariantly constants sections of $\LL^r$
($r$ rational).
One then writes a natural Lagrangian which yields all the field equations:
the Einstein equation and the equation for torsion;
the equation $F=2\,\dO Y$
(thus $Y$ is essentially the electromagnetic potential)
and the other Maxwell equation;
the Dirac equation~\cite{C00b}.

On the other hand, by fixing the tetrad $\Th$ and the gravitational part
of the spin connection one works in a fixed curved background structure.
This is the setting of this paper
and my previous articles about quantum theory.
Then $\Th$ allows the identification \hbox{$\TO\M\cong\LL\tn\H$},
and 1-forms of $\M$ can be viewed as \emph{scaled} sections
$\M\to\LL^{-1}\tn\H^*$.

\subsection{Gauge theories}
\label{ss:Gauge theories}

The two-spinor treatment of electrodynamics suggests a natural procedure
of generating gauge theories in some generality,
though not in \emph{all} generality:
we'll be able to recover the standard model and some possible extensions.
In particular, we do not aim at a theory
in which gravitation is on the same footing as other fields
(see~\sref{ss:Concluding remarks}).
Also note that, in our approach, the role of spinors is quite special,
not at all analogous to other internal degrees of freedom.

Our starting assumption is  that fermion fields can be described as sections
of a vector bundle \hbox{$\Y\onto\M$} where
$$\Y\equiv\YR\oplus\YL\equiv(\FR\tn\U)\oplus(\FL\tn\Ua)~,$$
and where $\F_{\!\!\Rrr}\onto\M$ and $\F_{\!\!\Lll}\onto\M$
are complex vector bundles,
describing the internal fermion structure besides spin,
endowed with fibered Hermitian structures
(fibered tensor products and direct sums over $\M$).
Next, by expanding $\Yc\tn\Y$, we'll notice that its sectors
are natural candidates for the role of boson bundles.
Further possible sectors may arise from the expansion of other tensor products.
Because of the algebraic structure of the fibers
one gets various contractions among different sectors,
which we view as related to the possible particle interactions.
Roughly speaking, the various tensor factors could be seen as an
analogue of ``chemical bonds''.

Explicitly, the expansion of our candidate boson bundle yields
\begin{align*}
\Yc\tn\Y&\cong
(\YRc\tn\YR)\oplus(\YLc\tn\YL)~\oplus~(\YRc\tn\YL)~\oplus~(\YLc\tn\YR)~\cong~{}
\\[6pt]
&\cong~
(\FRc\tn\FR\tn\Uc\tn\U)~\oplus~
(\FLc\tn\FL\tn\Ul\tn\Ua)~\oplus~{}\\[4pt]
&\qquad\qquad{}\oplus~(\FRc\tn\FL\tn\Uc\tn\Ua)~\oplus~
(\FLc\tn\FR\tn\Ul\tn\U)~.
\end{align*}
The Hermitian structures of $\FR$ and $\FL$ yield fibered isomorphisms
\hbox{$\FRc\cong\FRl$} and \hbox{$\FLc\cong\FLl$}\,;
we also have
\hbox{$\U\tn\Uc\cong\CC\tn\H$}, \hbox{$\Ul\tn\Ua\cong\CC\tn\H^*$}\,;
the Lorentz metric yields the isomorphism \hbox{$\H\leftrightarrow\H^*$},
and the tetrad $\Th$ yields the scaled isomorphism
\hbox{$\H^*\leftrightarrow\LL\tn\TS\M$}.
Hence, after rearranging the order of tensor factors,
sections \hbox{$\M\to\LL^{-1}\tn\YRc\tn\YR$ and $M\to\LL^{-1}\tn\YLc\tn\YL$}
can be seen as fields
\hbox{$\M\to\TS\M\tn\FR\tn\FRl$ and $\M\to\TS\M\tn\FL\tn\FLl$},
respectively, suitable for playing the role of gauge fields.

As for the last two terms in the above bundle expansion,
we note that the identity is a distinguished section of
\hbox{$\Ul\tn\U\cong\End\U$}.
A similar observation holds for \hbox{$\Uc\tn\Ua\cong\End\Uc$}.
If we restrict our consideration, in these sectors,
to sections which are proportional to the identity,
we obtain sections \hbox{$\M\to\FRc\tn\FL$} and \hbox{$\M\to\FLc\tn\FR$}\,,
suitable for the role of \emph{Higgs fields}.
On the other hand, nothing forbids to consider a larger class of fields;
that would complicate the matter considerably, but could be intriguing
in consideration of the still elusive properties
of the recently detected Higgs bosons.
Furthermore, one may examine the expansions of $\Y\tn\Yl$, $\Y\tn\Ya$
and their conjugate bundles.
Most sectors are, up to natural isomorphisms, the same already picked,
but we do get some new ones.
In particular, we get $\FR\tn\FRl$ and $\FL\tn\FLl$\,,
suitable for describing \emph{ghost fields}.
So, our scheme for generating a gauge theory,
though somewhat restricted with respect to full generality,
still leaves the room for various kinds of natural extensions.

Eventually, recalling that the Hermitian structures of $\FR$ and $\FL$
determine (\sref{ss:Classical gauge freedom}) Lie algebra sub-bundles
\hbox{$\Lie_\Rrr\subset\FR\tn\FRl$} and \hbox{$\Lie_\Lll\subset\FL\tn\FLl$},
we make the further assumptions that the targets of gauge and ghost fields
are restricted accordingly,
so that the field list in ``momentum representation'' is:
\smallbreak\noindent
$\bullet$~the \emph{matter field}
\hbox{$\Psi\equiv(\Psi_\Rrr\,,\,\Psi_\Lll):\P\to\Y$}\,,\quad $\P\equiv\TS\M$\,;

\smallbreak\noindent
$\bullet$~\emph{gauge fields}
\hbox{$W_{\Rrr}:\P\to\P\tn\Lie_\Rrr$} and
\hbox{$W_{\Lll}:\P\to\P\tn\Lie_\Lll$}\,;

\smallbreak\noindent
$\bullet$~\emph{Higgs and anti-Higgs fields},
\hbox{$\phi:\P\to\FL\tn\FRl$} and
\hbox{$\phi^\dag\equiv\bar\phi:\P\to\FR\tn\FLl$}\,;

\smallbreak\noindent
$\bullet$~\emph{ghosts} \hbox{$\ghR:\P\to\Lie_\Rrr$} and
\hbox{$\ghL:\P\to\Lie_\Lll$}\,;

\smallbreak\noindent
$\bullet$~\emph{anti-ghosts} \hbox{$\aghR:\P\to\Lie_\Rrr^*$} and
\hbox{$\aghL:\P\to\Lie_\Lll^*$}\,.

\smallbreak
We remark that ghosts and anti-ghosts are considered as independent fields,
though the geometric structure would allow a precise relation between them;
on the other hand, $\phi$ and $\bar\phi$ are mutually conjugated
(could be independent fields as well in some extended theory).

\smallbreak
In previous papers~\cite{C10a,C12a} I showed in some detail
how the above scheme fits electroweak theory,
with the settings \hbox{$\FL\equiv\I$} (the \emph{isospin bundle})
and \hbox{$\FR\equiv\weu2\I$}.

\subsection{Symmetry breaking}
\label{ss:Symmetry breaking}

Our general picture of a gauge theory has to be completed
by a description of symmetry breaking.
The ``vacuum value'' of the Higgs field is assumed to be a section
$$\Hivac:\M\to\FL\tn\FRl~,$$
which determines a splitting
$$\FL=\FR'\dir{\M}\FR^\sbot~,\quad \FR'\equiv\Hivac(\FR)~.$$
We'll assume $\Hivac$ to be of maximal rank $\dim\FR$\,,
so that it determines an isomorphism \hbox{$\FR\to\FR'\subset\FL$}\,.
Then the matter field can be decomposed as
$$\Psi\equiv(\Psi_\Rrr\,,\,\Psi_\Lll)
=(\Psi_\Rrr\,,\,\Psi'_\Rrr\,,\,\Psi_\Rrr^\sbot)\equiv(\psi\,,\,\n)~,$$
where
\begin{align*}
&\psi\equiv(\Psi_\Rrr\,,\,\Psi'_\Rrr):
\M\to (\FR\tn\U)\oplus(\FR'\tn\Ua)\cong\FR\tn\W~,
\\[6pt]
&\n\equiv\Psi_\Rrr^\sbot:\M\to\FR'\tn\Ua\subset\FL\tn\Ua~.
\end{align*}

The field $\Hivac$ can be regarded as an added feature
of the underlying classical geometric structure.
It's natural to assume it has the further property of being
conformally isometric, namely
$$h_\Lll\comp(\bHivac,\Hivac)=
\frac{\m^2}{\scriptstyle\dim\FR}\,h_\Rrr~,\quad \m\in\RR~,$$
where \hbox{$h_\Lll:\M\to\FLa\tn\FLl$}
and \hbox{$h_\Rrr:\M\to\FRa\tn\FRl$}
denote the Hermitian metrics of $\FR$ and $\FL$\,.
This condition implies \hbox{$\bang{\bHivac,\Hivac}=\m^2$},
so that $\Hivac$ is a minimum of the ``Higgs potential''
$$\l\,(2\,\m^2\,\bang{\bar\phi,\phi}-\bang{\bar\phi,\phi}^2)~,\quad
\l\in\RR^+\,,$$
where \hbox{$\bang{\bar\phi,\phi}\equiv
\bang{h_\Lll\comp(\bar\phi,\phi),h_\Rrr^\#}$}
denotes the scalar obtained by contraction of \hbox{$\bar\phi\tn\phi$}
via the Hermitian structure.

The $\Hivac$-splitting of $\FL$\,, together with the metric $h_\Lll$\,,
yields a splitting \hbox{$\FRl=\FR'{}^\lin\oplus\FR^{\sbot\lin}$},
so that
\begin{align*}
\Lie_\Lll\subset\End\FL&=
(\FR'\oplus\FR^\sbot)\tn(\FR'{}^\lin\oplus\FR^{\sbot\lin})=
\\[6pt]
&=(\FR'\tn\FR'{}^\lin)
\oplus(\FR^\sbot\tn\FR'^\lin)
\oplus(\FR'\tn\FR^{\sbot\lin})
\oplus(\FR^\sbot\tn\FR^{\sbot\lin})~.
\end{align*} 
Now consider the decomposition of any \hbox{$\xi\in\Lie_\Lll$} as
$$\xi=\xi'+\xi^{+}+\xi^{-}+\xi^\sbot
\in\Lie_\Rrr'\oplus\Lie_\Rrr^{+}\oplus\Lie_\Rrr^{-}\oplus\Lie_\Rrr^\sbot~,$$
where \hbox{$\Lie_\Rrr'\subset\FR'\tn\FR'^\lin$},
\hbox{$\Lie_\Rrr^\sbot\subset\FR^\sbot\tn\FR^{\sbot\lin}$}\,, and
$$\Lie_\Rrr^{+}\equiv\FR^\sbot\tn\FR'^\lin~,\quad
\Lie_\Rrr^{-}\equiv\FR'\tn\FR^{\sbot\lin}~.$$
Since $\Lie_\Rrr^{+}$ and $\Lie_\Rrr^{-}$ are anti-isomorphic
by Hermitian adjunction,
and any \hbox{$\xi\in\Lie_\Lll$} (being anti-Hermitian)
fulfills \hbox{$(\xi^{-})^\dag=-\xi^{+}$},
eventually we get a splitting
$$\Lie_\Lll\cong
\Lie_\Rrr'\oplus\Lie_\Rrr^{+}\oplus\Lie_\Rrr^\sbot\cong
\Lie_\Rrr'\oplus\Lie_\Rrr^{-}\oplus\Lie_\Rrr^\sbot~,$$
and, accordingly, a decomposition of the gauge, ghost and anti-ghost fields
in the left sector.

\section{Quantum states and interactions}
\label{s:Quantum states and interactions}
\subsection{Quantum states as generalised semi-densities}
\label{ss:Quantum states as generalised semi-densities}

Let \hbox{$\Z\onto\X$} be a finite-dimensional complex vector bundle,
\hbox{$\dim_\RRr\X=m$}\,.
Assume that $\X$ is \emph{orientable}, and choose a positive
semi-vector bundle \hbox{$\VV\equiv\VV\X\equiv(\weu{m}\TO\X)^+$}\,.
A section \hbox{$\X\to\VV^{-1/2}\tn\Z$} is called
a \emph{$\Z$-valued semi-density}.
The vector space of all such sections which are smooth and have compact support
is denoted as $\DCho(\X,\Z)$\,.
The dual space of $\DCho(\X,\Zl)$ in the standard topology~\cite{Sc}
is indicated as \hbox{$\DCh(\X,\Z)$}
and called the space of \emph{generalised semi-densities} of \hbox{$\Z\onto\X$}.
In particular, a sufficiently regular ordinary section
\hbox{$\th:\X\to\VV^{-1/2}\tn\Z$} can be seen as an element in $\DCh(\X,\Z)$
via the rule
\hbox{$\bang{\th,\s}:=\int_\X\bang{\th(x),\s(x)}$}\,,
\hbox{$\s\in\DCho(\X,\Zl)$}\,.

Semi-densities have a special status among all kinds of generalised sections,
since there is a natural inclusion $\DCho(\X,\Z)\subset\DCh(\X,\Z)$\,.
Furthermore, if a fibered Hermitian structure
of \hbox{$\Z\onto\X$} is assigned then one has the space $\LC^2(\X,\Z)$
of all ordinary semi-densities $\th$ such that
\hbox{$\bang{\th^\dag,\th}<\infty$}\,.
The quotient \hbox{$\HC(\X,\Z)=\LC^2(\X,\Z)/\boldsymbol{0}$}
is then a Hilbert space
(here \hbox{$\boldsymbol{0}\subset\LC^2(\X,\Z)$} denotes the subspace
of all almost-everywhere vanishing sections),
and we get a so-called \emph{rigged Hilbert space}~\cite{BLT}
$$\DCho(\X,\Z)\subset\HC(\X,\Z)\subset\DCh(\X,\Z)~.$$
Elements in \hbox{$\DCh(\X,\Z)\setminus\HC(\X,\Z)$} can then be identified
with the (\emph{non-normalisable}) \emph{generalised states}
of the common terminology.

Let $\d[x]$ be the \emph{Dirac density} on $\X$
with support $\{x\}$\,, \hbox{$x\in\X$}\,.
A generalised semi-density is said to be \emph{of Dirac type}
if it is of the form \hbox{$\d[x]\tn v\in\DCh(\X,\Z)$}
with \hbox{$v:\X\to\VV^{1/2}\tn\Z$}.
If $\bigl(\bb_\a\bigr)$ is a frame of \hbox{$\Z\onto\X$}
then we set
$$\ket{x}\tn\bb_\a(x) \leftrightarrow
\Bsf_{x,\a}\equiv\d[x]\tn\eta^{-1/2}\tn\bb_\a(x)~,$$
and call the set \hbox{$\bigl(\Bsf_{x,\a}\bigr)\subset\DCh^1$}
a \emph{generalised basis}.
Accordingly we introduce a handy ``generalised index'' notation.
We write \hbox{$\Bsf^{x,\a}\equiv\d[x]\tn\eta^{-1/2}\tn\bb^\a(x)$}\,,
where $\bigl(\bb^\a\bigr)$ is the dual classical frame.
Though contraction of two distributions is not defined in general,
a straightforward extension of the discrete-space operation yields
$$\bang{\Bsf^{x'\!,\a'},\Bsf_{x,\a}}=\d^{x'\!,\a'}_{x,\a}\,\eta(x)~.$$
which is consistent with ``index summation'' in a generalised sense:
if \hbox{$f\in\DCho(\X,\Z)$}
and \hbox{$\l\in\DCho(\X,\Zl)$} are test semi-densities,
then we write
\begin{align*}
&f^{x,\a}\equiv f^\a(x)\equiv\bang{\Bsf^{x,\a},f}~,\quad
\l_{x,\a}\equiv \l_\a(x)\equiv\bang{\l,\Bsf_{x,\a}}~,
\\[6pt]
&\bang{\l,f}\equiv
\l_{x'\!,\a'}\,f^{x,\a}\,\bang{\Bsf^{x'\!,\a'},\Bsf_{x,\a}}\equiv
\int_\X \l_\a(x)\,f^\a(x)\,\eta(x)~,
\end{align*}
namely we interpret index summation
with respect to the continuous variable $x$ as integration
(provided by the chosen volume form).
This formalism can be extended to the contraction
of two generalised semi-densities whenever it makes sense.

Next we set \hbox{$\ZC^1\equiv\DCh(\X,\Z)$}\,,
which is our template for the space of states of one particle of some type.
The associated ``$n$-particle state'' space $\ZC^{n}$ is defined
to be either the symmetrised tensor product $\vee^n\ZC^1$ (\emph{bosons})
or the  anti-symmetrised tensor product $\weu{n}\ZC^1$ (\emph{fermions}).
The ``multi-particle state'' space is defined to be
\hbox{$\ZC\equiv\bigoplus_{n=0}^\infty\ZC^{n}$}
(constituted by finite sums with arbitrarily many terms).
Similarly we set \hbox{$\ZC^{\lin1}\equiv\DCh(\X,\Zl)$} and define
the ``dual'' multi-particle space to be
\hbox{$\ZC^\lin\equiv\bigoplus_{n=0}^\infty\ZC^{\lin n}$}.

A general theory of quantum particles has several particle types.
Correspondingly, one considers several multi-particle state spaces
(or ``sectors'') $\ZC'$, $\ZC''$, $\ZC'''$ etc.
The total state space is now defined to be
$$\VC:=\ZC'\tn\ZC''\tn\ZC'''\tn\mdots=
\textstyle{\bigoplus_{n=0}^\infty}\VC^{n}$$
where $\VC^n$, constituted of all elements of tensor rank $n$\,,
is the space of all states on $n$ particles of any type.
We observe that if $\XC$ and $\YC$ are any two vector spaces,
then their antisymmetric tensor algebras fulfill the isomorphisms
$$\weu{p}(\XC\,{\oplus}\,\YC)\cong
\textstyle{\bigoplus_{h=0}^p}\,(\weu{p-h}\XC)\tn(\weu{h}\YC)~,\quad
(\wedge\XC)\tn(\wedge\YC)\cong\wedge(\XC\,{\oplus}\,\YC)~.$$
Hence all fermionic sectors can be described by a unique
overall antisymmetrised tensor algebra.
A similar observation holds true for the bosonic sectors,
while we regard mutual ordering of fermionic and bosonic sectors as inessential.
Similarly one constructs a ``dual'' space
\hbox{$\VC^\lin:=\ZC^\lin{}'\tn\ZC^\lin{}''\tn\ZC^\lin{}'''\mdots=
\bigoplus_{n=0}^\infty\VC^{\lin n}$}.

\subsection{Quantum bundles, detectors and free-particle states}
\label{ss:Quantum bundles, detectors and free-particle states}

In this paper we use the term \emph{quantum bundle} to mean
a vector bundle over spacetime whose fibers
are distributional spaces~\cite{C05,C12a}.
The underlying ``classical'' (\ie\ finite-dimensional)
geometric structure is that of a 2-fibered bundle,
and the infinite-dimensional smooth structure is conveniently treated
in terms of Fr\"olicher's notion of smoothness,
or \emph{F-smoothness}~\cite{Fr,FK,KM,JM02,CK95,MK}.
The F-smooth geometry of distributional bundles and quantum connections
has been studied in a previous paper~\cite{C04a}.

Let $(\M,g)$ be Einstein's spacetime.
Taking the speed of light and the Planck constant into account,
the covariant form of a particle's 4-momentum
is valued into \hbox{$\Pm\subset\P\cong\TS\M$},
the sub-bundle over $\M$ of future `mass -shells'
corresponding to mass $m\in\{0\}\cup\LL^{-1}$
($\LL$ is the semi-space of length units).
Let now $\Z\to\Pm$ be a vector bundle
(representing the `internal degrees of freedom'
of the considered particle type).
The constructions of~\sref{ss:Quantum states as generalised semi-densities}
at each \hbox{$x\in\M$},
with the generic manifold $\X$ replaced by $(\Pm)_x$\,,
yield spaces $\ZC^1_x$\,,
and the fibered set \hbox{$\ZC^1:=\bigsqcup_{x\in\M}\!\!\ZC^1_x$}
turns out to have a natural F-smooth vector-bundle structure over $\M$.
The multi-particle bundle
\hbox{$\ZC:=\bigoplus_{n}\ZC^{n}\onto\M$},
\hbox{$n\in\{0\}\cup\NN$}\,,
can also be straightforwardly constructed.

Let $g^\#$ be the ``contravariant'' metric
induced on the fibers of \hbox{$\P\equiv\TS\M$}.
When an observer (a congruence of timelike curves) is considered
then one has the orthogonal splitting
\hbox{$\P=\P_{\!\!\spar}\oplus_{{}_\T}\P_{\!\!\sbot}$}\,,
and the volume form $\eta_\sbot$
on the fibers of \hbox{$\P_{\!\!\sbot}\onto\M$}.
The orthogonal projection
\hbox{$\P\to\P_{\!\!\sbot}$} yields a distinguished diffeomorphism
\hbox{$\Pm\leftrightarrow\P_{\!\!\sbot}$} for each $m$.
The pull-back of $\eta_\sbot$ is then
a volume form on the fibers of $\Pm$\,,
which is denoted for simplicity by the same symbol.

It will be convenient to to use the ``spatial part''
$p_\sbo$ of the 4-momentum $p$ as a label,
that is a generalised index for quantum states.
For each \hbox{$p\in\Pm$} let $\d_m[p]$
the Dirac density with support $\{p\}$ on the same fiber of $\Pm$\,,
and \hbox{$\d(\yy_\sbo\!{-}p_\sbo)$} the generalised function characterised by
\hbox{$\d_m[p](\yy)=\d(\yy_\sbo\!{-}p_\sbo)\,\dO^3\yy$}\,,
where we are using linear coordinates
\hbox{$\bigl(\yy_\l\bigr)\equiv\bigl(\yy_0,\yy_1,\yy_2,\yy_3\bigr)
\equiv\bigl(\yy_0,\yy_\sbo\bigr)$}
in the fibers of $\P$.
Now consider the section \hbox{$\Pm\to\DCh(\Pm,\CC):p\mapsto\Xsf_p$}
defined as follows;
for each \hbox{$p\in\Pm$} we can regard $\Xsf_p$
as a generalised function of the variable $\yy_\sbo$\,,
with the expression
$$\Xsf_p(\yy):=l^{-3/2}\,\d(\yy_\sbo{-}p_\sbo)\,\sqrt{\dO^3\yy}~.$$
Here $l$ is a constant length needed
in order to get an unscaled semi-density
(compare with the usual ``box quantization'' argument).
Eventually, we get the distinguished isomorphism
\hbox{$\ZC^1\leftrightarrow\uDCh(\Pm,\Z)$}
which is determined by the correspondence
\hbox{$\ket{z}\leftrightarrow\Xsf_p\tn z$}\,, \hbox{$z\in\Z_p$}\,.

We can develop our arguments
by assuming a weaker structure than a congruence of spacetime submanifolds,
namely a unique timelike submanifold \hbox{$\T\subset\M$},
which we call a \emph{detector}.
Through a natural construction exploiting the exponential map,
we also get a timelike congruence in a neighbourhood of $\T$.
Actually this setting suffices for reproducing,
in terms of generalised semi-densities,
essentially the standard momentum space formalism,
which can be seen as a sort of a complicated `clock'
carried by the detector~\cite{C05,C12a}.

We obtain a \emph{generalised frame of free one-particle states}
along $\T$ as follows.
First, at some arbitrarily fixed event \hbox{$t_0\in\T\subset\M$}
we choose a frame \hbox{$\bigl(\bb_\a\bigr)$} of \hbox{$\Z\to(\Pm)_{t_0}$}\,.
Then the family of generalised semi-densities
\hbox{$\Bsf_{p\a}(t_0)\equiv\Xsf_p\tn\bb_\a\in\DCh(\Pm,\Z)_{t_0}$}
is a generalised frame at $t_0$\,.
We transport $\Bsf_{p\a}$ along $\T$
by means of Fermi transport~\cite{C09a,C12a}
for the spacetime and spinor factors,\footnote{
Fermi and parallel transport coincide if the detector is inertial.
} 
and, for the remaining factors,
by means of parallel transport relatively to a suitable connection of $\Z$
which will have to be assumed (see also~\sref{ss:Concluding remarks}).
We write
$$\Bsf_{p\a}:\T\to\DCh(\Pm,\Z)_\Tt:
t\mapsto\Bsf_{p\a}(t)=\Xsf_{p(t)}\tn\bb_\a~,$$
where \hbox{$p:\T\to\Pm:t\mapsto p(t)$} is Fermi-transported.
This yields a trivialization
$$\DCh(\Pm,\Z)_\Tt\cong\T\times\DCh(\Pm,\Z)_{t_0}~,$$
which can be seen as determined by a suitable connection called
the \emph{free-particle connection}.
Eventually, the above arguments can be naturally extended
to multi-particle bundles and states.
When several particle types are considered,
we get a trivialization \hbox{$\VC_{\!\Tt}\cong\T\times\QC$}
of the total quantum state bundle
so that $\QC\equiv\VC_{\!t_0}$ can be identified with the space 
of all asymptotical quantum states.
The quantum interaction can be constructed,
assembling the classical interaction with a distinguished quantum ingredient,
as a modification of that parallel transport.
The free-particle trivialization
preserves particle type and number by construction,
while the interaction doesn't.

\subsection{Quantum interactions}
\label{ss:Quantum interactions}

Quantum interactions are described by a section
$$-\iO\,\dt\tn\Hfr:\T\to\TO^*\T\tn\End(\VC)~,$$
where the scaled function $\tt$ is the detector's \emph{proper time}.
A \emph{quantum history} is defined to be a section \hbox{$\T\to\VC_{\!\Tt}$}\,,
which we conveniently regard as a map
\hbox{$\psi:\T\to\QC\equiv\VC_{\!t_0}$}\,, obeying the law
\hbox{$\psi(t)=\Ucal_{t_0}(t)\,\psi(t_0)$}\,,
where \hbox{$\Ucal_{t_0}:\T\to\End(\VC_{\!t_0})$}
is determined by the differential equation
$$\frac{\dO}{\dO t}\Ucal_{t_0}(t)=-\iO\,\Hfr(t)\comp\Ucal_{t_0}(t)~,\qquad
~\Ucal_{t_0}(t_0)=\Id{\QC}~.$$
The free-particle connection
yields the trivialisation \hbox{$\VC_{\!\Tt}\to\T\times\QC$}\,;
we can see the interaction as a tensor field
which modifies that connection and determines a new quantum connection
of the functional bundle \hbox{$\VC_{\!\Tt}\onto\T$}.
Or, the interaction can be seen as a 1-form on $\T$
valued into the endomorphisms of the fixed space $\QC$\,.

The time-dependent endomorphism $\Hfr$ is dictated,
in an essentially elementary way,
by the underlying ``classical structure'',
while the problem of determining $\Ucal$ is on a different footing.
In perturbative theories one starts from the \emph{Dyson series},
which provides a formal solution of the above differential equation,
and tries to extract meaningful physical results from the study
of the \emph{scattering operator}
$$\Scal:=
\lim_{\begin{subarray}~~t\,\to\,{+\infty} \\ t_0\to\,{-\infty} \end{subarray}}
\Ucal_{t_0}(t)~,$$
which, intuitively, relates asymptotical states of `incoming'
and `outgoing' free particles interacting in a small spacetime region.

Essentially, $\Hfr$ arises as the tensor product of the classical interaction
and a certain semi-density on particle momenta
(the ``quantum ingredient'' of the interaction).
Consider masses $m',m'',\dots,m^{\sst{(r)}}$ and let the shorthand
$$\Pmm:=\Px{m'}\cart{\M}\Px{m''}\cart{\M}\mdots\cart{\M}\Px{m^{\sst{(r)}}}
\onto\M$$
denote the bundle of $r$ particle momenta corresponding to these masses.
Let $\der$ be the fiberwise generalised function\footnote{
Namely $\der$ is a section of a distributional bundle~\cite{C04a} over $\M$.
For the sake of brevity, in this paper we are not going to explicitly denote
all involved spaces.} 
on $\Pmm$ characterised by
$$\bang{\der\,,\,f}=
\iint\breve f\bigl(\yy'_\sbo\,,\yy''_\sbo\,,\dots\,,\yy^{\sst{(r-1)}}_\sbo\,,
-{\textstyle\sum_{i=1}^{r-1}\yy^{\sst{(i)}}_\sbo}\bigr)\,
\dO^3\yy'\,\dO^3\yy''\,\dots\,\dO^3\yy^{\sst{(r-1)}}$$
for any test density
\hbox{$f=\breve f\,\dO^3\yy'\tn\dO^3\yy''\tn\mdots\tn\dO^3\yy^{\sst{(r)}}$},
namely
\hbox{$\der\equiv\d(\yy'_\sbo{+}\yy''_\sbo{+}\dots{+}\yy^{\sst{(r)}}_\sbo)$}
in standard notation.
Recalling that on the fibers of \hbox{$\Pm\onto\M$}
we have the natural \emph{Leray form},
which can be then written as
\hbox{$\om_m(p)=(2\,p_0)^{-1}\eta_\sbot(p)$}\,, \hbox{$p\in\Pm$}\,,
where \hbox{$p_0\equiv\Eo_m(p_\sbo)=(m^2+p_\sbo^2)^{1/2}$},
we introduce the generalized half-density
$$\uLar:=\der\,\sqrt{\om_{m'}}\tn\mdots\tn\sqrt{\om_{m^{\sst{(r)}}}}=
\frac{\d(\yy'_\sbo{+}\mdots{+}\yy^{\sst{(r)}}_\sbo)}%
{\sqrt{2^{{}^{\scriptstyle r}}\yy_0'\dots\yy_0^{\sst{(r)}}}}\,
\sqrt{\dO^3\yy'}\tn\dots\tn\sqrt{\dO^3\yy^{\sst{(r)}}}~.$$
By multiplying $\uLar$ by certain phase factors
we introduce a modified generalized half-density $\QIr$,
which can be expressed in the generalised index notation as\footnote{
Generalised index summation is interpretd as integration
(\sref{ss:Quantum states as generalised semi-densities})
via the volume form $\eta_\sbot$\,.} 
\begin{align*}
&\QIr=\QI^{p'p''p'''\!\!\dots{}}\:
\Xsf_{p'}\tn\Xsf_{p''}\tn\Xsf_{p'''}\tn\mdots{}~,
\\[6pt]
&\QI^{p'p''p'''\!\!\dots{}}\:\equiv\:
l^{2\,r}\,(2^r\,p_0'\,p_0''\,p_0'''\mdots{})^{\sst-1/2}\,
\eO^{-\iO\,(p_0'+p_0''+p_0'''+\mdots)\,\tt}\,
\d(p'_\sbo+p''_\sbo+p'''_\sbo+\mdots)~.
\end{align*}
We consider a special rule for lowering and rising indices in $\QIr$,
so obtaining tensors of different index types, associated with $\QIr$,
as in finite-dimensional tensor algebra.
This rule (which can be seen as related to a Hermitian structure)
prescribes that moving an index $p^{\sst(i)}$
you change the sign of $p_0^{\sst(i)}$ in the exponential
and the sign of $p^{\sst(i)}_\sbo$ in the delta-function, so that
$$\QI_{p'}{}^{p''p'''\!\!\dots{}}\: = \:
l^{2\,r}\,\frac{\eO^{-\iO\,(-p_0'+p_0''+p_0'''+\mdots)\,\tt}}%
{(2^r\,p_0'\,p_0''\,p_0'''\mdots{})^{\sst1/2}}\,
\d(-p'_\sbo+p''_\sbo+p'''_\sbo+\mdots)$$
and the like.

Now the tensor field $\ellr$
describing the classical interaction of $r$ particles
(\ie\ the interaction lagrangian)
yields as many index types as $\QIr$,
and the corresponding index types from $\ellr\tn\QIr$
generate all the various pieces of the quantum interaction,
namely the various terms in $\Hfr$\,.
A term with $s$ low indices describes,
via a generalised analogue of an elementary algebraic mechanism,
the absorption of $s$ particles and the emission of $r\,{-}\,s$\,.
Propagators and all particle interactions in gauge theories
can be indeed recovered from the above ideas
(see previous papers~\cite{C05,C12a} for details).

\section{Gauge freedom}
\label{s:Gauge freedom}
\subsection{Classical gauge freedom}
\label{ss:Classical gauge freedom}

If the ``matter field'' of a classical theory
is a section of the bundle \hbox{$\E\onto\M$},
then the classical ``gauge field'' is a connection of that bundle,
namely a section \hbox{$\E\to\JO\E$} of the 1-jet bundle.
If \hbox{$\E\onto\M$} is a vector bundle then, in particular,
we consider linear connections,
which can be seen as sections \hbox{$\M\to\GA$} where
\hbox{$\GA\subset\JO\E\ten{\M}\E^*$}
is the sub-bundle projecting over the identity $\Id{\E}$.
This is an affine bundle, with ``derived'' vector bundle
\hbox{$\DO\GA=\TS\M\ten{\M}\End\E$}
(the bundle of ``differences of linear connections'').

The fibers of the vector bundle \hbox{$\End\E\onto\M$}
are constituted by all linear endomorphisms of the respective fibers of $\E$,
and are naturally Lie algebras via by the ordinary commutator.
In fact, this is the Lie algebra bundle of the group bundle
\hbox{$\Aut\E\onto\M$} of all automorphisms of the fibers of $\E$.
However, \hbox{$\E\onto\M$} is usually endowed
with some fibered geometric structure,
which selects the (``internal'' symmetry)
Lie-group subbundle \hbox{$\Gb\onto\M$}
of all automorphisms preserving it;
the fibers of $\Gb$ are isomorphic Lie groups,
though distinguished isomorphisms among them
don't exist in general.\footnote{
In order to deal with a \emph{fixed group}
one can exploit the notion of a principal bundle.} 
The Lie algebra bundle of $\Gb$ is a sub-bundle \hbox{$\Lie\subset\End\E$}.
By ordinary restriction we obtain the affine sub-bundle
\hbox{$\GA_{{\!}_\Gb}\subset\GA$}\,, with derived vector bundle
\hbox{$\DO\GA_{{\!}_\Gb}=\TS\M\ten{\M}\Lie$}\,;
sections \hbox{$\M\to\GA_{{\!}_\Gb}$} characterise those linear connections
which preserve the fiber geometric structure
(\ie\ make it covariantly constant).
Hence the difference of any two such connections is $\Lie$-valued.

Connections can be locally described as tensor fields
by choosing a gauge, namely a local ``flat'' connection $\g_0$\,.
In fact the difference
\hbox{$\a\equiv\g\,{-}\,\g_0:\E\to\TS\M\ten{\E}\VO\E$}
determines an arbitrary connection $\g$\,, and,
since we are dealing with linear connections, we can write
$$\a:\M\to\TS\M\ten{\M}\End\E\equiv\TS\M\ten{\M}\E\ten{\M}\E^*~.$$
The curvature tensor \hbox{$\r\equiv[\g,\g]$} (Fr\"olicher-Nijenhuis bracket)
can be expressed in terms of $\a$ as \hbox{$2\,[\g_0,\a]+[\a,\a]$}.

The $\g_0$-constant local sections of \hbox{$\E\onto\M$}
determine a trivialization of $\E$ over any sufficiently small
open subset of $\M$.
Thus one also has $\g_0$-constant local frames.
Conversely, the assignment of a local frame determines a flat connection $\g_0$
by the condition that its coefficients vanish in that frame.

A (local) \emph{gauge transformation} is defined to be
a section \hbox{$K:\M\to\Gb$}\,.
Together with its transposed inverse $\cev{K}{}^*$,
a fibered automorphism of \hbox{$\E^*\onto\M$},
it determines a fibered automorphism of the whole tensor algebra
of $\E\cart{\M}\E^*$.
Moreover, $K$
transforms the family of $\g_0$-constant sections to a new family of sections,
which determines a new flat connection
\hbox{$\g_0'=\g_0+(\nabla[\g_0]K)\pint\cev K$}\,.
In particular, if $\nabla[\g_0]K=0$ then $\g_0'=\g_0$\,,
namely the two families 
of covariantly constant sections coincide.\footnote{
In that case one uses to say
that $K$ is a ``global'' gauge transformation.} 

Let now \hbox{$\g=\a-\g_0=\a'-\g_0'$} be a fixed connection;
we get \hbox{$\a'-\a=(\nabla[\g_0]K)\pint\cev K$},
namely \hbox{$\a,\a':\M\to\TS\M\ten{\M}\Lie$} represent the same connection
whenever their difference is of that type.
This implies that any scalars formed
from covariant derivatives of tensor fields
and from the curvature tensor of $\g$ are invariant under the replacement
\hbox{$\a\leftrightarrow\a'$}.
So we can look at the notion of gauge freedom as follows:
if we insist in describing gauge fields
(\ie\ connections) in terms of tensor fields
then we concede them too many ``degrees of freedom'',
which must be absorbed by taking a suitable quotient.
The crucial point is that in quantum theory the fields
must be sections of vector bundles.

Still in view of quantum theory we consider gauge fields in terms of momenta.
We take a hint from the observation that a radiative electromagnetic field
is usually assumed~\cite{Li60} to be of the form \hbox{$F=k\we b$}\,,
with \hbox{$k,b:\M\to\TS\M$} such that $k^\#$ is a geodesic null vector field
and \hbox{$g^\#(k,b)=0$}\,.
While in curved spacetime we may not be able to find a \emph{closed}
such tensor field,\footnote{
Radiative e.m.\ fields in curved spacetime are usually dealt with
by considering solutions of the Maxwell equations which approximate said type
in the small wavelength limit.} 
it makes sense to use it as a template for our description of photons.
The couple $(k,b)$ constitutes then a \emph{redundant} description,
being a representative of an equivalence class
characterising the e.m.\ potential.
We can describe a more general gauge field by a couple $(k,\a)$\,,
with \hbox{$\a:\M\to\TS\M\ten{\M}\Lie$}\,,
such that \hbox{$k^\#\pint\a=0$}\,.
The physical meaning of the gauge field is encoded by its equivalence class,
$(k,\a)$ and $(k,\a')$ being equivalent
if their difference is of the kind $k\tn\chi$ with \hbox{$\chi:\M\to\Lie$}\,.

The equivalence class of $(k,\a)$ also uniquely determines the
``curvature-like'' tensor
$$\r[k,\a]:=\iO\,k\we\a+\a\bwe\a~,$$
where the notation $\a\bwe\b$
stands for exterior product of $\Lie$-valued forms
together with composition.\footnote{
In components,
\hbox{$(\a\bwe\b)\iI{ab}i=\sco ijk\a\iI aj\,\b\iI bk$} where
\hbox{$\sco ijk\equiv[\lfr_j\,,\,\lfr_k]^i$}
are the ``structure constants'' in the chosen frame
$\bigl(\lfr_i\bigr)$ of $\Lie$.} 
In the quantum theory in momentum representation,
the Lagrangian for the gauge field, which is expressed in terms of $\r$\,,
yields all self-interaction terms.
The replacement \hbox{$\a\to k\tn\chi+\a$}
does not affect any scattering matrix calculations.
According to the usual quantisation procedure,
this freedom can be exploited by adding to the Lagrangian density
a suitable term (proportional to the squared divergence of $\a$)
which is not gauge-invariant,
namely does not ``pass to the quotient'' when we
deal with the above said equivalence classes,
though it is a natural geometric object when $\a$ is seen as a tensor field.
This modifies the gauge particle propagator in a way that
does not affect point interactions.

\subsection{Two-spinors and one-particle states in QED}
\label{ss:Two-spinors and one-particle states in QED}

Let \hbox{$\Pm\onto\M$}
(\sref{ss:Quantum bundles, detectors and free-particle states})
be the sub-bundle of $\TS\M$
whose fibers are the mass-shells
corresponding to mass \hbox{$m\in\{0\}\cup\LL^{-1}$}.
If \hbox{$p\in(\Pm)_x$}\,, \hbox{$x\in\M$},
then we have the \emph{Dirac splitting}
$$\W\!\!_x=\W^{+}_{\!\!p}\oplus\W^{-}_{\!\!p}~,\quad
\W^{\pm}_{\!\!p}:=\Ker(\g[p^\#]\mp m)~,$$
where \hbox{$p^\#\equiv g^\#(p)\in\LL^{-2}\tn\TO\M$}
is the contravariant form of $p$\,.
Thus we obtain 2-fibered bundles \hbox{$\W_{\!\!m}^\pm\onto\Pm\onto\M$}, where
$$\W_{\!\!m}^\pm:=\bigsqcup_{p\in\Pm}\W_{\!\!p}^\pm\subset\Pm\cart{\M}\W~.$$
We call $\W^{+}_{\!\!m}$ and $\Wc{}^{-}_{\!\!m}$
the \emph{electron bundle} and the \emph{positron bundle}, respectively.
If $\bigl(\zeA(p)\bigr)$ is a 2-spinor frame such that
$p^\#\propto\t_{\!_0}$ in the associated Pauli frame,
then the \emph{Dirac frame}
$\bigl(\uu_\sA(p)\,,\vv_\sB(p)\,\bigr)$\,, defined by
$$\uu_1\equiv\osq\,(\z_1\,,\bze^1)~,~~
\uu_2\equiv\osq\,(\z_2\,,\bze^2)~,~~
\vv_1\equiv\osq\,(\z_1\,,-\bze^1)~,~~
\vv_2\equiv\osq\,(\z_2\,,-\bze^2)~,$$
is $\kO$-orthonormal and adapted to the Dirac splitting.

The splitting has an interesting two-spinor description~\cite{C07}.
If \hbox{$\psi\equiv(u,\bl)\in\W$} then
$$\t\equiv\tfrac1{\sqrt2\,|\bang{\l,u}|}\,(u\tn\bu+\l^\#\tn\bl^\#)\in\H$$
is a unit future-pointing timelike vector.
By a straightforward calculation one sees that \hbox{$\g[\t]\psi=\pm\psi$}
if and only if \hbox{$\bang{\l,u}\in\RR^\pm$}.
Conversely, it can be proved that if \hbox{$\t'\in\H$} is such that
\hbox{$\g[\t']\psi=\pm\psi$}\,,
then necessarily \hbox{$\t'=\t$}\,.
In other terms, \emph{internal states of free electrons and positrons
carry the full information about their momenta}.

For a fixed a detector \hbox{$\T\subset\M$}, 
we use generalised electron and positron frames
\begin{align*}
& \Asf_{p\sA}:=\eO^{-\iO\,p_0\,t}\,\Xsf_p\tn\uu\!_\sA(p)
:\T\to\DCh(\Pm\,,\W_{\!\!m}^+)~,\\[6pt]
& \Csf_{p\cA}:=\eO^{-\iO\,p_0\,t}\,\Xsf_p\tn\bvv\!_\cA(p)
:\T\to\DCh(\Pm\,,\Wc_{\!\!m}^-)~,
\end{align*}
where \hbox{$p:\T\to\Pm$} is Fermi-transported.

We discuss real photon states in spacetime terms first.
Consider the zero-mass subbbundle \hbox{$\Pz\subset\TS\M$}
of future null half-cones.
We use the identification \hbox{$\H^*\cong\LL\tn\TS\M$}
determined by the fixed tetrad.
Let \hbox{$\H'\subset\Pz\cart{\M}\H^*$} be the sub-bundle over $\M$
whose fiber over any \hbox{$k\in(\Pz)_x$}\,, \hbox{$x\in\M$},
is the 3-dimensional real vector space
\hbox{$\H'_k:=\{y\in\H^* : g^\#(k,y)=0\}$}\,.
We have the real vector bundle \hbox{$\B_{\!\RRr}\onto\Pz$}
whose fiber over any \hbox{$k\in\Pz$}
is the 2-dimensional quotient space \hbox{$\H'_k/k$}\,.
It turns out that the (contravariant) spacetime metric `passes to the quotient',
so it naturally determines
a negative metric $g_\Bb$ in the fibers of \hbox{$\B_{\!\RRr}\onto\Pz$}\,,
as well as a `Hodge' isomorphism ${*}_{\!\Bb}$
which can be characterised through the rule
\hbox{${*}(k\we\b)=-k\we({*}_{\!\Bb}\b)$}\,.

The complexified 2-fibered bundle \hbox{$\B:=\CC\tn\B_{\!\RRr}\onto\Pz\onto\M$}
(the \emph{optical bundle}~\cite{C00b,Nu}) has the natural splitting
$$\B=\B^+ \dir{\Pz} \B^-~,$$
where the fibers of \hbox{$\B^\pm\to\Pz$}
are complex 1-dimensional $g_\Bb$-null subspaces defined to be
the eigenspaces of $-\iO\,{*}_{\!\Bb}$ with eigenvalues $\pm1$
(\emph{self-dual} and \emph{anti-self-dual} subspaces).
Restricting these bundles to the detector's world line \hbox{$\T\subset\M$}
then we can identify \hbox{$\B_\RRr\onto\Pz\onto\T$} with
\hbox{$\H'\cap\H^{*\sbot}\onto\Pz\onto\T$} (`radiation gauge').
For any $k\in(\Pz)_x$\,, $x\in\M$,
let $\bigl(\t^\l\bigr)$ be a Pauli basis of $\H$ at $x$
such that $\t_0$ is tangent to $\T$ and $k^\#\propto\t_0+\t_3$\,;
setting
\begin{align*}
&\bigl(\bb_{\sst+}\,,\bb_{\sst-})\equiv\bigl(\bb_1\,,\bb_2)
:=\Bigl(\osq\,(\t^1+\iO\,\t^2)~,~\osq\,(\t^1-\iO\,\t^2)\Bigr)
\subset\CC\tn\H'\cap\H^{*\sbot}~,
\\[6pt]
&\Bsf_{k\sQ}:=\eO^{-\iO\,k_0\,t}\,\Xsf_k\tn\bb_\sQ(k)~,
\quad k\in\Pz\,,~{\scriptstyle Q}=1,2\,,
\end{align*}
one gets, by Fermi transport,
a generalised frame $\bigl\{\Bsf_{\k\sQ}\bigr\}$ of the quantum bundle
$\DCh_{\!\Mm}(\Pz\,,\B)\onto\M$.
This frame is adapted to the self-dual/anti-self-dual splitting.

\smallbreak\noindent{\bf Remark.}~%
While the photon's  physical meaning is encoded in the 2-form $k\we\b$
(\sref{ss:Classical gauge freedom}),
the radiation gauge determines $k$ and $\b$ separately.
\smallbreak

We now observe (\sref{ss:Two-spinor basics})
that an element \hbox{$\b\in\CC\tn\H^*=\Ul\tn\Ua$}
is null if and only if it is decomposable (\ie\ a monomial),
while future-pointing real null elements are of the type \hbox{$k=\k\tn\bk$}\,.
It's not difficult to check that \hbox{$\b\in\B_{\!k}^\pm$}
iff \hbox{$\b=\k\tn\bl$} and \hbox{$\b=\l\tn\bk$}\,, respectively.
On the other hand, sums of the type \hbox{$\k\tn\bl+\m\tn\bar\n$}
span the whole $\CC\tn\H^*$,
which is also spanned by \emph{virtual photons}.
For the latter we can enlarge the frame $\bigl(\bb_{\sst+}\,,\bb_{\sst-})$
by including, for example,
\hbox{$\bb_0\equiv\t^0$} and \hbox{$\bb_3\equiv\osq\,(\t^0-\t^3)$}\,,
respectively related to ``scalar'' and ``longitudinal'' modes.

\subsection{QED interactions and gauge freedom in terms of two-spinors}
\label{ss:QED interactions and gauge freedom in terms of two-spinors}

In electrodynamics, the algebraic part of the point interaction
can be described as the tensor field
\hbox{$\ell\interaction:\M\to\Wa\tn\H\tn\Wl$} defined by
$$\ell\interaction(\bar\phi,A,\psi):=-e\,\bang{\bar\phi,\g[A^\#]\psi}~,$$
where $e$ is the positron's charge.
Index moving in the fibers is determined by Dirac adjunction\footnote{
We do \emph{not} consider different index positions
obtained through some positive Hermitian metric:
that would be an extra structure,
equivalent to the assignment of an observer.} 
in $\W$ and by the Lorentz metric in $\H$.
By using each factor in $\ell\interaction$ either as absorption or as emission
we obtain the eight point interactions of QED, represented by the diagrams
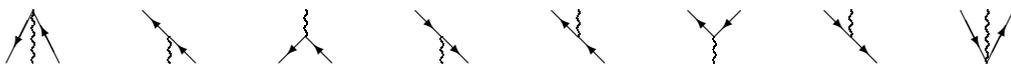
\begin{align*}
&\Lcqs 
&&\LSqs 
&&\LcQs 
&&\LCqc 
&&\LSQs 
&&\LSqC 
&&\LCQc 
&&\LCQS 
\end{align*}
(time flows upwards).
In two-spinor terms, if
\hbox{$\phi=(v,\bm)$} and \hbox{$\psi=(u,\bl)$} we get
$$\bang{\bar\phi,\g[r\tn\bs]\psi}=
\sqrt2\,\bigl(\bang{\m,r}\,\bang{\bl,\bs}+\e(u,r)\,\be(\bv,\bs)\bigr)\equiv
\sqrt2\,g\bigl(r\tn\bs\;,\;u\tn\bv+\m^\#\tn\bl^\#\bigr)~,$$
hence in general
\hbox{$\bang{\bar\phi,\g[A^\#]\psi}=
\sqrt2\,g\bigl(A^\#,u\tn\bv+\m^\#\tn\bl^\#\bigr)$}\,.
The kernel of the map
$$\bang{\bar\phi,\g[\_]\psi}:
\CC\tn\H^*\to\CC:A\mapsto\bang{\bar\phi,\g[A^\#]\psi}$$
is then constituted by all covectors orthogonal to
\hbox{$u\tn\bv+\m^\#\tn\bl^\#\in\U\tn\Uc=\CC\tn\H$}.

In particular we observe that setting
$$k\equiv\frac{m}{\surd2}\Bigl(\frac{(u\tn\bu)^\fl+\l\tn\bl}{|\bang{\l,u}|}
\pm\frac{(v\tn\bv)^\fl+\m\tn\bm}{|\bang{\m,v}|}\Bigr)$$
by straightforward 2-spinor algebra calculations one obtains
$$\tfrac1m\,\bang{\bar\phi,\g[k^\#]\psi}=
\bang{\m,u}\,
\Bigl(\frac{\bang{\bl,\bu}}{|\bang{\l,u}|}
\pm\frac{\bang{\bm,\bv}}{|\bang{\m,v}|}\Bigr)
+\bang{\bl,\bv}\,
\Bigl(\frac{\bang{\l,u}}{|\bang{\l,u}|}
\pm\frac{\bang{\m,v}}{|\bang{\m,v}|}\Bigr)~.$$
It's easy to check~\cite{C07}
that the condition \hbox{$\psi\equiv(u,\bl)\in\W^{\pm}$}
can be expressed, in 2-spinor terms, as \hbox{$\bang{\l,u}\in\RR^{\pm}$}.
Hence the above expression vanishes when the minus sign applies
and $\phi$ and $\psi$ represent internal spaces
of either two electrons or two positrons,
and also vanishes when the plus sign applies
and we are dealing with mutual antiparticles.
Moreover if $\phi$ and $\psi$ represent free fermions
then $k$ is either the sum or the difference of their momenta,
so that the above depicted point interactions 
are unaffected by adding the algebraic sum
of the interacting fermions' momenta to the internal photon state.

Now consider virtual fermions connecting to a node in a Feynman diagram.
A fermion's propagator contains a factor \hbox{$\id\pm\frac1m\,\g[p]$}
for an electron (resp.\ positron) of momentum $p$\,,
and we must bear in mind that $p$ is now an integration variable
spanning the whole $\P$.
Now if $p$ is ``off-shell'' then \hbox{$\id\pm\frac1m\,\g[p]$}
is \emph{not} a projection onto $\W^\pm$.
However we recall that the covariant propagator,
in which the dependence on time has been eliminated,
is actually the sum of two contributions,
corresponding to different time ordering of the propagator's nodes.
If we keep the time-dependent description,
in which particles only ``travel forward'' in time,
then a  fermion's propagator contains a factor
$$\id\pm\frac1m\,\g[\Eo_m(p_\sbo)+p_\sbo]\equiv
\id\pm\frac1m\,\g[p_0+p_\sbo]~,$$
where the plus is for electrons and the minus is for positrons.
This is the projection onto $\W_p^\pm$
with \hbox{$p\equiv\Eo_m(p_\sbo)+p_\sbo\in\Pm$}\,,
hence the above arguments can be extended to this situation.

Finally, we note that these results can be straightforwardly extended
to more general gauge theories
of the type described in~\sref{ss:Gauge theories}.

\subsection{Concluding remarks}
\label{ss:Concluding remarks}

In quantum theory all fields, including gauge fields,
must be sections of some vector bundle.
This requirement can be understood at least from two different points of view.
In the ``momentum representation'',
as sketched in~\sref{ss:Quantum bundles, detectors and free-particle states},
the construction of the distributional bundle describing quantum states
requires a vector bundle over particle momenta,
whose fibers describe the particle's ``internal states''.
In the ``position representation'' the fields are (generalized) sections
of a vector bundle \hbox{$\OC\tn\E\onto\M$},
where \hbox{$\E\onto\M$} is the classical configuration bundle
(a vector bundle whose sections are the fields
of the theory under consideration)
and $\OC$ is an infinite-dimensional $\ZZ_2$-graded algebra,
generated by absorption and emission operators.\footnote{
The relation between these two approaches in terms of F-smooth geometry
will be examined in a forthcoming paper.} 

According to a third, quite different point a view~\cite{Pe71,C11a},
\emph{the system defines the geometry} and reality is fundamentally discrete;
any notion related to continuity should be recovered as a convenience
in the description of sufficiently complex systems.
Ideas of this kind have been around for some time
and have inspired a few tries at serious theories~%
\cite{Rovelli2010,Verlinde2010,Sorkin2003},
but, as far as I know, no definitely convincing results have been obtained.
In Loop Quantum Gravity, in particular,
certain discrete geometric structures are the basic quantum states,
but how ordinary matter enters the scheme is still unclear.
By contrast I propose that physical reality \emph{is} fundamentally a network,
whose nodes and edges we call \emph{events} and \emph{particles}, respectively.
In a sufficiently large portion of the network,
approximate geometric relations will emerge among its external edges;
on the other hand, knowing about some external edges
we can guess at other external edges in probabilistic terms.
So spacetime, gravity (\emph{not} quantum gravity) and quantum mechanics
could all emerge form a more fundamental discrete theory.

The relation between spin and spacetime geometry supports these ideas.
Rather than trying to recover Euclidean geometry from general spin networks,
we could focus our attention on networks whose edges are labeled
by internal states,
taking the relations among spin and particle's momentum
(\sref{ss:Two-spinors and one-particle states in QED})
into account.
A possible way of undertaking this task is to try and immerse
these networks into a manifold with suitable properties,
to be chosen in such a way to allow us to derive experimentally
testable consequences.
Since a measure apparatus is macroscopic,
such consequences must be of statistical nature.

Spacetime metric and bundle connections belong to the macrosopic notions
which allow us to handle the physics;
in this sense they could be viewed as ``mean field'' background properties
of a physical system.
Gauge particles, in particular, are related to connections,
as the relation must consider the partial indeterminacy
of the particles' internal states
when expressed in terms of spacetime geometry.



\end{document}